\documentclass[twocolumn,twoside,slac_two]{revtex4}
\usepackage{graphicx}
\usepackage{fancyhdr}
\usepackage{graphics}
\usepackage{epstopdf}
\usepackage{textpos}
\newcommand{\pp}{\mbox{p--p}}
\newcommand{\PbPb}{\mbox{Pb--Pb}}
\newcommand{\sqrts}{\sqrt{s}}
\newcommand{\sqrtsNN}{\sqrt{s_{\mathrm{NN}}}}
\newcommand{\lsim}{\,{\buildrel < \over {_\sim}}\,}

\newcommand{\av}[1]{\left\langle #1 \right\rangle}

\newcommand{\gev}{\mathrm{GeV}}
\newcommand{\tev}{\mathrm{TeV}}

\newcommand{\mub}{\mathrm{\mu b}}

\newcommand{\pt}{p_{\rm t}}
\newcommand{\Et}{E_{\rm t}}
\newcommand{\ptave}{\langle p_{\rm t} \rangle}

\newcommand{\dNchdeta}{{\rm d}N_{ch}/{\rm d}\eta}
\newcommand{\Npart}{N_{\rm part}}
\newcommand{\Ncoll}{N_{\rm coll}}

\newcommand{\Dzero}{{\rm D^0}}

\newcommand{\Dplus}{{\rm D^+}}
\newcommand{\jpsi}{{\rm J/}\psi}

\newcommand{\Raa}{R_{\rm AA}}
\newcommand{\missptave} {\langle \displaystyle{\not} p_{\mathrm{T}}^{\parallel} \rangle} %%%
\begin{document}

%%%%%%%%%%%%%%%%%%%%%% WRITE THE TITLE HERE %%%%%%%%%%%%%%%%%%%
\title{\centering Heavy-ion results from the LHC}
%%%%%%%%%%%%%%%%%%%%%% WRITE THE AUTHOR HERE %%%%%%%%%%%%%%%%%
\author{
\centering
\includegraphics[scale=0.15]{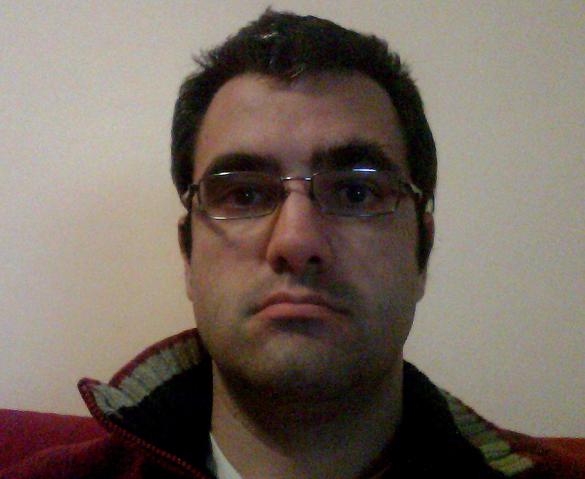} \\
\begin{center}
F. Prino
\end{center}}
\affiliation{\centering \mbox{INFN, Sezione di Torino, Italy}}
%%%%%%%%%%%%%%%%%%%%%% WRITE THE ABSTRACT HERE %%%%%%%%%%%%%%%%
\begin{abstract}

\end{abstract}

\maketitle
\thispagestyle{fancy}


%%
\section{Introduction}
\label{sec:intro}

In November 2010, the first heavy-ion run began at the LHC.
Pb nuclei were collided at the centre-of-mass energy $\sqrtsNN=2.76~\tev$, 
about 14 times higher than that achieved at RHIC.
The integrated luminosity delivered by the LHC during the 5 weeks of running
was 10~$\mub^{-1}$.
Three experiments collected data with $\PbPb$ collisions, namely 
ALICE, ATLAS and CMS.

Heavy-ion collisions at relativistic energies are aimed at studying nuclear 
matter at extreme conditions of temperature and energy density, where Lattice 
QCD predicts the matter to be in a state where quarks and gluons are 
deconfined over volumes much larger than the size of a hadron 
(see e.g.~\cite{Lattice}).
Such a state is called Quark Gluon Plasma (QGP).
The goal of heavy-ion collision experiments is to collect 
evidence for the existence of this new state of matter and to study its 
properties.

$\PbPb$ collisions at the LHC are expected to generate a
medium that has a higher initial temperature and energy density that that
generated at lower values of $\sqrts$.
Experimental measurements in this new energy regime are a key benchmark for
models that reproduce the features observed at lower collision energy.
Furthermore, the input from the LHC is also crucial to shed light on some 
issues that are not completely understood from the SPS and RHIC results (e.g. 
the $\jpsi$ suppression).
Also, the steep increase with $\sqrts$ of the cross-section for QCD 
scatterings with high virtuality results in copious production of hard partons  
at the LHC, thus allowing one to reach higher precision on the 
experimental observables related to high momentum and heavy 
flavoured particles.
Finally, at LHC energies, new (higher mass) probes, such as W and Z$^0$ 
bosons, become available and provide new tools to study the properties of the 
medium.

\section{Collision geometry and system evolution}
\label{sec:evol}

Nuclear collisions are characterized by the impact parameter between the
colliding nuclei, i.e. the distance between the centers of the nuclei in the 
transverse plane.
The impact parameter defines the centrality of the collision: in head-on
(``central'') collisions a large number of nucleons participates in the
interaction and many binary nucleon-nucleon collision occur, leading to a 
large energy deposit in the reaction volume.
Conversely, collisions with large impact parameter (``peripheral'') have 
lower number of participant nucleons and binary collisions and consequently 
lower energy densities are attained.
Experimentally, the definition of the collision centrality is based on
the assumption that the impact parameter is monotonically related 
to the particle multiplicity or the energy produced in the collision.
It is typically expressed in terms of fractions (percentiles) of the 
total hadronic cross section.
Starting from the centrality-related experimental observables,
the collision geometry can be reconstructed utilizing the Glauber 
model~\cite{glauber1}, which describes a nuclear collision in terms of 
multiple-scattering of nucleons in nuclear targets, together with a model 
for particle production.
This approach allows one to compute the average numbers of participant nucleons 
($\langle \Npart \rangle$), of elementary nucleon-nucleon collisions 
($\langle \Ncoll \rangle$), and other geometrical quantities for each 
centrality interval~\cite{glauber2}.

The main difference between a nucleus-nucleus collision and a \pp\ collision
is the larger collision volume where a dense medium of 
strongly interacting matter (fireball) is formed.
The evolution of the produced medium goes through various stages which are 
schematically described in the following (see e.g.~\cite{HeinzConcepts} for
more details).
The fireball first goes through an early non-equilibrium stage.
Re-scatterings among its constituents can establish thermal equilibrium.
If the system thermalizes quickly enough and at sufficiently large energy 
density, it is expected to be in the QGP phase.
Experimental results indicate that, for central Au-Au collisions at RHIC, 
thermal equilibrium is reached at a time $\tau_{eq}=0.6~$fm/$c$ after the 
collision when the temperature of the fireball is $\approx$ 360 
MeV, well above the critical values for deconfinement~\cite{HeinzConcepts}.
The thermalized medium undergoes a collective hydrodynamic expansion driven
by the pressure gradients generated by the high temperatures and densities 
that characterize these stages of the collision evolution~\cite{hydro}.
As a consequence, the fireball cools down and its energy density decreases.
When the energy density reaches the critical threshold for deconfinement
($\epsilon \sim 1~\gev/{\rm fm}^3$), the partons convert to hadrons 
(hadronization).
After this phase transition, hadrons keep re-scattering off each other, thus
building up further collective flow, until the system becomes so dilute that 
all hadronic interactions cease and the particles decouple 
(``freeze-out'').
The freeze-out is actually anticipated to occur in two 
stages~\cite{HeinzConcepts}.
First (at higher temperatures) inelastic processes stop, thus
freezing the abundances of the various hadron species (chemical freeze-out).
Afterwards (at lower temperatures), also elastic collision among hadrons cease
and the particle momenta get frozen (thermal freeze-out).
At the thermal freeze-out, the bulk of hadrons which underwent the 
thermalization and the successive collective expansion have an approximately
exponential transverse momentum spectrum reflecting the temperature of the 
fireball at that point, blue-shifted by the average transverse collective 
flow~\cite{HeinzConcepts}.
The evolution of the fireball from the moment when local thermal 
equilibrium is reached until the thermal freeze-out can be described with
hydrodynamics~\cite{hydro}.

The bulk of particles produced in heavy-ion collisions are low-momentum 
(soft) hadrons that have undergone the collective expansion and 
have abundances and momenta that are fixed at the stages of the chemical 
and thermal freeze-out.
Experimental observables for low-momentum particles produced in Au-Au 
collisions at RHIC are well described by hydrodynamics with a QGP equation 
of state and very low shear viscosity (see e.g.~\cite{hydro,hydro2}).
The results from the LHC in the soft particle sector are described in 
Section~\ref{sec:soft}.

On top of the bulk of soft particles, also hard particles, with either a large
mass or large transverse momenta ($\pt \gg 1~\gev/c$), are created.
Hard particle production occurs in the early stages of the collision, before 
the bulk of quanta have time to re-scatter and thermalize.
Once produced, hard particles have to traverse the medium constituted
by the bulk of soft particles and can therefore be used as probes for the 
medium properties as discussed in Section~\ref{sec:hard}.

\section{Global event characteristics}
\label{sec:soft}

\begin{figure}[tb!]
\centering
\includegraphics[width=0.42\textwidth]{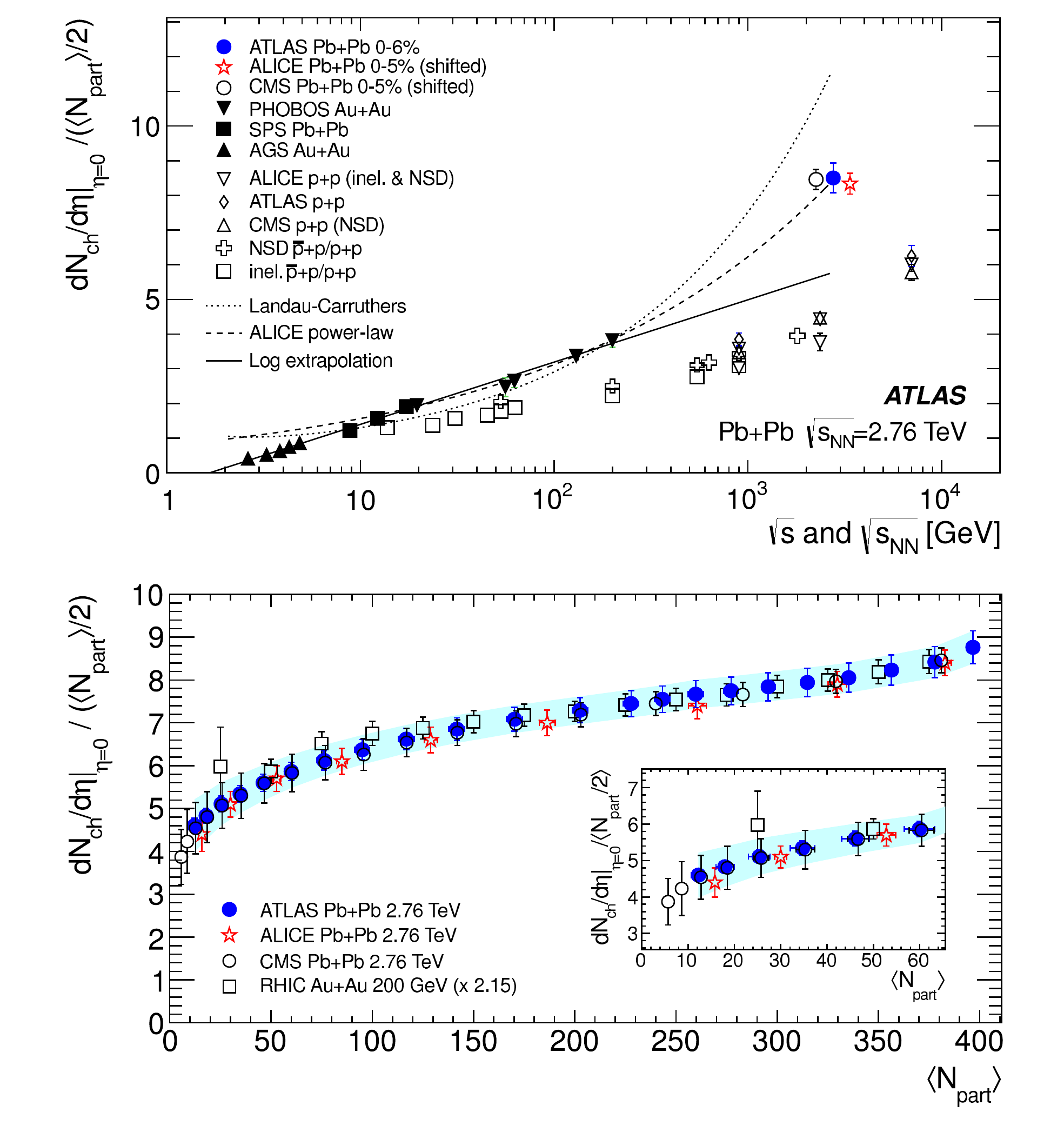}
\caption{Top: Collision energy dependence of $\dNchdeta$ per colliding 
nucleon pair in $\pp$ (open symbols) and AA (closed symbols) collisions. 
Bottom: ($\dNchdeta$)/($\Npart$/2) vs. centrality at LHC compared
to RHIC results scaled by a factor 2.15.
Taken from~\cite{ATLASmult}.}
\label{fig:dndeta}
\end{figure}

The measurement of the multiplicity of produced particles,  
quantified by the charged particle density per unit of rapidity ($\dNchdeta$), 
is a global observable that provides insight into the density of 
gluons in the initial stages and on the mechanisms of particle production.
All three experiments have measured $\dNchdeta$ at mid-rapidity as a function
of the collision centrality~\cite{ALICEmult1, ALICEmult2, ATLASmult, CMSmult}.
The multiplicity in the most central collisions at the LHC is 
larger by a factor $\approx$2.1 with respect to central collisions at top 
RHIC energy, in disagreement with the scaling with log$\sqrts$ observed at 
lower energies (see top panel of Fig.~\ref{fig:dndeta}).
The centrality dependence of ($\dNchdeta$)/($\Npart$/2) has a similar shape 
to that observed at RHIC (Fig.~\ref{fig:dndeta}, bottom panel)
and is reasonably reproduced both by models based on
gluon saturation in the initial state and by two-component Monte Carlo 
models~\cite{ALICEmult2,CMSmult}.

The produced transverse energy $\Et$ was estimated by ALICE by measuring 
the charged hadronic energy with the tracking system and adding the 
contribution of neutral particles.
The measured $\Et$ per (pseudo)rapidity unit can be used to estimate the 
energy density with the Bjorken formula~\cite{Bjorken}:
\begin{equation}
\label{Bjorken}
\varepsilon_{Bj}=\frac{1}{\mathcal{A} \tau} \left. \frac{d\Et}{dy}\right|_{y=0}
\end{equation}
where $\mathcal{A}$ is the transverse overlapping area in the collision of the 
two nuclei and $\tau$ is the formation time.
For the most central (0--5\%) collisions at the LHC, the resulting value is 
$\varepsilon_{Bj}\tau \approx 16~\gev/({\rm fm}^2c)$, about a factor
3 larger than the corresponding value at RHIC~\cite{ToiaQM,CMSdetdeta}.

The system size is measured from the HBT radii extracted from the study of 
two-pion correlations.
For central collisions, it is found to be larger 
by a factor two with respect to the one observed in central collisions at 
the top RHIC energy~\cite{ALICEhbt}.

A typical feature of the medium produced in heavy-ion collisions is the 
presence of collective motions arising from the large pressure gradients
generated by compressing and heating the nuclear matter.
The first type of collective motion is called radial flow and it stems from 
the isotropic expansion of the fireball.
It is accessed experimentally by measuring the transverse momentum ($\pt$) 
spectra of identified hadrons which at low $\pt$ show a thermal (Boltzmann)
distribution blue-shifted by the collective velocity of the system expansion.
The $\pt$ spectra of identified hadrons ($\pi$, K and p) have been measured
by ALICE in a wide momentum range~\cite{FlorisQM}.
The spectra are seen to be harder (i.e. characterized by a less steep 
distribution and a larger $\ptave$) than those observed at RHIC at 
$\sqrtsNN$=200~$\gev$.
This is a first indication for a stronger radial flow at the LHC.
Deeper insight is obtained by fitting the $\pi$, K and p spectra with a 
blast-wave function~\cite{BlastWave}, which allows one to extract an estimate 
of the parameters of the system at the thermal freeze-out, namely the 
freeze-out temperature and the radial flow velocity. 
The radial flow velocity from blast-wave fits results, for the most central 
collisions at the LHC, about 10\% higher than what observed in central
collisions at top RHIC energy~\cite{FlorisQM}.

The build-up of a collective motion is also signaled by the presence of 
anisotropic flow patterns in the transverse plane due to an initial 
geometrical anisotropy in the spatial distribution of the nucleons 
participating in the collision~\cite{Ollitrault}.
Rescatterings among the produced particles convert this initial geometrical 
anisotropy into an observable momentum anisotropy.
For non-central collisions, the geometrical overlap region of the colliding 
nuclei present an almond-like shape and the impact parameter defines a 
preferred direction in the transverse plane.
The anisotropy of produced particles is characterized by the Fourier 
coefficients $v_n=\langle\cos[n(\varphi-\Psi_n)]\rangle$,
where $n$ is the order of the harmonic, $\varphi$ is the azimuthal angle of the
particle and $\Psi_n$ is the angle of the initial state spatial plane of 
symmetry.
If the matter distribution in the colliding nuclei varied smoothly, the plane 
of symmetry would coincide with the reaction plane (i.e. the plane defined by 
the impact parameter and the beam direction) and odd Fourier coefficients 
would be zero by symmetry. 
However, due to fluctuations in the spatial distribution of the participating
nucleons in the colliding nuclei, the plane of symmetry $\Psi_n$ fluctuates 
event-by-event relative to the reaction plane. 
This gives rise to the presence of odd harmonics, such as $v_3$ and $v_5$, in
the particle azimuthal distributions.
Since the impact parameter vector and the spatial planes of symmetry 
are not directly measurable, various experimental techniques have been
developed to estimate the anisotropic flow coefficients from measured 
correlations among the observed particles~\cite{PosVolSne}.

\begin{figure}[tb!]
\centering
\includegraphics[width=0.42\textwidth]{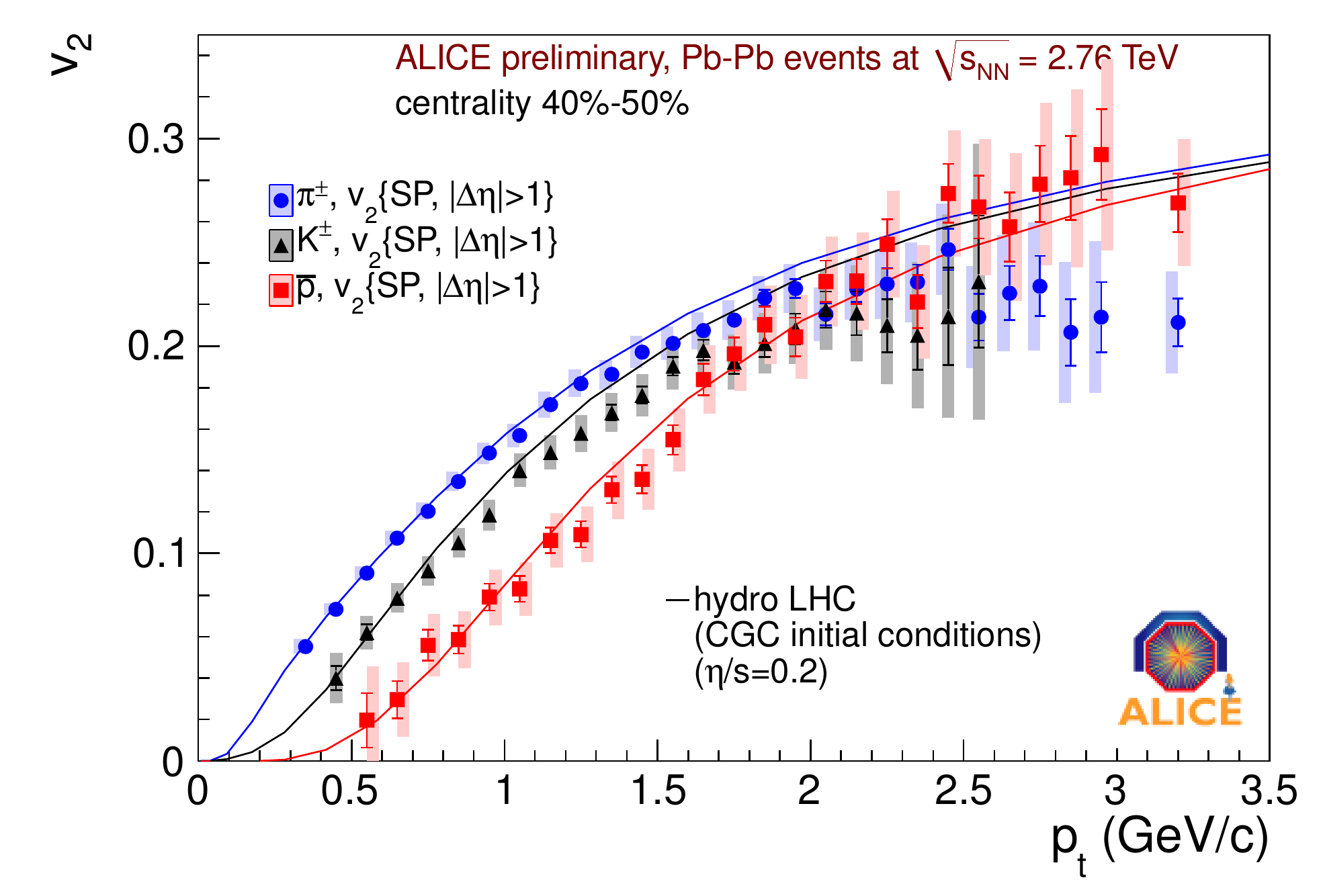}
\caption{Measured elliptic flow of identified particles for centrality 
bin 40\%-50\%, compared to hydro predictions from~\cite{Heinz}}
\label{fig:v2pKpi}
\end{figure}

Elliptic flow has been studied by all three 
experiments~\cite{ALICEflow,ATLASflow,CMSflow}.
The $\pt$ integrated elliptic flow of charged particles is found to
increase by about 30\% from the highest RHIC energy of $\sqrts=200~\gev$ to
LHC energy~\cite{ALICEflow}.

The $\pt$-differential elliptic flow is sensitive to the evolution and
freeze-out conditions of the expanding medium. 
The measured $v_2(\pt)$ at the LHC is found to be compatible with that
observed at RHIC.
The 30\% increase in the integrated elliptic flow is therefore coming from
an increase in the average transverse momentum, due to the increase of 
the radial flow with increasing $\sqrts$.
The larger radial flow leads also to a more pronounced mass dependence
of the elliptic flow. 
In Fig.~\ref{fig:v2pKpi} the measured $v_2(\pt)$ for identified pions, kaons 
and protons is shown for semi-peripheral (40--50\%) collisions.
Hydrodynamic model predictions, based on the assumption that the QGP shear 
viscosity over entropy ratio does not change from RHIC to LHC~\cite{Heinz}, 
are shown in the figure and provide a good description of $v_2(\pt)$ for the 
three particle species. 
This is not the case in more central (10--20\%) reactions where
the hydrodynamic prediction does not provide a reasonably good description of 
anti-protons~\cite{SnellingsQM}.
This mismatch is also observed for the $\pt$ spectra of identified 
protons~\cite{FlorisQM} and it may be due
to a larger radial flow in the data. 
At RHIC energies, a better description of the antiproton flow was 
obtained by introducing a hadronic cascade after the hydrodynamic evolution.
The results obtained for elliptic flow indicate that the hot and dense matter 
created in heavy-ion collisions at the LHC still behaves like 
a strongly interacting fluid with exceptionally low viscosity, as that 
observed at RHIC~\cite{hydro2,Heinz}.

Higher Fourier harmonics have been measured by the three 
experiments~\cite{ALICEhighharm,ATLAShighharm,CMShighharm}.
In the most central events, dominated by fluctuations in the initial 
geometrical configuration, the elliptic ($v_2$) and triangular
($v_3$) flow are found to have similar magnitude.

The presence of higher harmonics provides a natural explanation of the 
structures observed in the two-particle azimuthal correlations at low $\pt$, 
namely the ``ridge'' at $\Delta\varphi\approx0$ and large $\Delta\eta$ and the 
double-bump in the away side ($\Delta\varphi\sim\pi$). 
These structures were first observed at RHIC in Au-Au 
collisions~\cite{RHICridge} and their interpretation is a long standing puzzle:
models based on jet-medium interactions were at first developed 
(e.g. ~\cite{Mach}), while more recently an explanation based on triangular 
flow was proposed~\cite{AlverRoland}.
These structures in the two-particle correlations have been studied in 
detail by performing a Fourier analysis of the $\Delta\eta - \Delta\varphi$ 
correlations with large $\Delta \eta$ 
gap~\cite{ATLAShighharm,ALICE2part,CMS2part}.
The data are found to be well described by the first five terms of the Fourier 
series.
At low $\pt$ (i.e. $\pt \lsim 3-4~\gev/c$), the Fourier components 
extracted from two-particle correlations are found to factorize
into single-particle harmonic coefficients which are in agreement
with the anisotropic flow coefficients.
This suggests that the features observed in two-particle correlations at low 
$\pt$ are consistent with the collective response of the system to the 
initial state geometrical anisotropy.

\section{Characterization of the medium with hard probes}
\label{sec:hard}

Particles with large transverse momentum and/or mass, which are produced in 
large-virtuality parton scatterings, are powerful tools to probe the medium
created in heavy-ion collisions.
The production of such ``hard probes'' in nuclear collisions is expected 
to scale with the number of nucleon--nucleon collisions in the 
nucleus--nucleus collision (binary scaling).
The experimental observable used to verify the binary scaling is the nuclear 
modification factor:
\begin{equation}
\label{eq:Raa0}
R_{\rm AA}=\frac{\mathrm{Yield~in~AA}}
{\av{\Ncoll} \cdot \mathrm{Yield~in~pp}}
\end{equation}
where $\av{\Ncoll}$ is the average number of binary (nucleon-nucleon) 
collisions for the considered centrality class.
If no nuclear effects are present, $\Raa$ should be equal to 1 by construction.
It is anticipated that the medium created in the collision affects the 
abundances and spectra of the originally produced hard probes, resulting in
a value of $\Raa$ different from 1.
In particular:
\begin{itemize}
\item{Quarkonia are expected to melt in the QGP due to color charge
screening which would lead to a suppression of the measured yield of 
charmonia and bottomonia~\cite{MatsuiSatz}.}
\item{Partons are expected to lose energy while traversing the 
strongly-interacting medium, via gluon radiation and elastic collisions with 
the partonic constituents (see e.g.~\cite{eloss1,eloss2} for recent reviews).}
\end{itemize}
The $\Raa$ observable is therefore sensitive to the properties of the medium
created in the collision.
It has however to be considered that other effects related to the presence of 
nuclei in the initial state (e.g. nuclear modifications of the PDFs, Cronin 
enhancement) can break the expected binary scaling.

\subsection{Quarkonia}
\label{sec:quarkonia}

As mentioned above, quarkonium states are expected to be suppressed ($\Raa<1$)
in the QGP, due to the color screening of the force which binds the $c\bar{c}$
(or $b\bar{b}$) state~\cite{MatsuiSatz}.
The suppression is predicted to occur both for the charmonium ($\jpsi$, 
$\Psi^\prime$, $\chi_c$, ...) and bottomonium families 
(~$\Upsilon$(1S, 2S, 3S)~) above 
(or close to) the critical temperature for the phase transition.
Forthermore, the quarkonium suppression is anticipated to occur sequentially 
according to the binding energy, i.e. strongly bound states (such as $\jpsi$ 
and $\Upsilon$(1S)) should melt at higher temperatures relative to 
more loosely bound states.
It has also to be considered that with increasing $\sqrts$, due to the
more abundant production of charm in the initial state, charmonium regeneration 
from $c$ and $\bar{c}$ recombination at hadronization time can lead to an 
enhancement in the number of observed $\jpsi$~\cite{PBM}.

At the LHC, the $\jpsi$ yield has been measured by all the three experiments.
ATLAS measured $\jpsi$ at mid-rapidity and high $\pt$ (80\% of $\jpsi$'s measured
by ATLAS have $\pt>6.5~\gev/c$) and reported a suppression which increases
with increasing centrality~\cite{ATLASjpsi}.
CMS measured the nuclear modification factor of prompt and secondary $\jpsi$'s
at mid-rapidity, finding a suppression which increases with 
centrality~\cite{CMSquarkonia}.
Also in this case, high $\pt$ ($>6.5~\gev/c$) $\jpsi$ are measured.
The $\Raa$ measured at the LHC at mid-rapidity shows more suppression than 
observed by the PHENIX and STAR experiments at RHIC at central rapidity.
It should be noticed that PHENIX measured $\jpsi$ mesons down to $\pt=0$, 
while the preliminary results from STAR~\cite{STARqm} are for 
high-$\pt$ ($>5~\gev/c$) $\jpsi$'s and show a systematically higher $\Raa$ 
than observed by PHENIX at low-$\pt$.
It has also to be considered that the CMS results are for prompt $\jpsi$
at high $\pt$, while PHENIX and STAR measured inclusive $\jpsi$ 
(i.e. including those from B feed-down).
CMS also reported the $\Raa$ for non-prompt $\jpsi$'s 
from B feed-down which are found to be less suppressed than the prompt ones. 
In this case, the mechanism that gives rise to the observed suppression is 
not the quarkonium melting: B mesons decay well outside the medium, but
arise from the fragmentation of a b quark which suffers energy loss when 
traversing the colored medium.
Additional information about $\jpsi$ suppression at the LHC is presented 
by the ALICE measurement~\cite{GinesQM} which is performed at forward rapidity
and down to $\pt=0$ for inclusive $\jpsi$.
The resulting $\Raa$ shows a suppression almost independent of centrality and
smaller than that observed by PHENIX at forward rapidity.
In summary, the LHC results on $\jpsi$ suppression provide hints that the 
$\jpsi$ suppression is $\pt$ dependent and that regeneration may play an 
important role at low $\pt$.
A deeper understanding requires studies of initial state effects
by measuring $\jpsi$ production in p--A collisions at LHC energy.

CMS also measured $\Upsilon$ states in both $\pp$ and $\PbPb$ collisions
at $\sqrts=2.76~\tev$, resolving the 1S, 2S and 3S states.
The strongly-bound $\Upsilon(1S)$ state is found to be suppressed: the 
preliminary measurement of the nuclear modification factor integrated over 
centrality is $\Raa=0.62\pm0.11\pm0.10$~\cite{CMSquarkonia}.
The suppression of the higher-mass states is measured relative to the ground 
state, by building a double ratio between $\Upsilon$(2S+3S) and $\Upsilon$(1S)
 yields in $\PbPb$ and $\pp$.
The result is:
$$\frac{[\Upsilon(2S+3S)/\Upsilon(1S)]_{PbPb}}
{[\Upsilon(2S+3S)/\Upsilon(1S)]_{pp}}=0.31_{-0.15}^{+0.19}(stat)\pm0.03(syst)$$
which indicates that the excited states are significantly more suppressed
than the ground state.
The probability to measure such a value, or a lower one, if the true double 
ratio is equal to one, is less than 1\%~\cite{CMSupsilon}.

\subsection{Parton energy loss in the QCD medium}

As mentioned above, hard partons, produced at the initial stage of the 
collision, traverse the medium and are expected to be sensitive to its 
energy density, through the mechanism of in-medium partonic energy 
loss.
The amount of energy lost is sensitive to the medium properties (density)
and depends also on the path-length of the parton in the deconfined matter
as well as on the properties of the parton probing the medium.
The nuclear modification factor as a function of $\pt$ is therefore 
considered:
$$R_{\rm AA}(\pt)=
{1\over \av{N_{\rm coll}}}  
{d N_{\rm AA}/d\pt \over 
d N_{\rm pp}/d\pt} = 
{1\over \av{T_{\rm AA}}}  
{d N_{\rm AA}/d\pt \over 
d\sigma_{\rm pp}/d\pt}$$
where $\av{T_{\rm AA}}$ is the average nuclear overlap function  computed with the Glauber model for the considered centrality class~\cite{glauber2}.
The parton energy loss manifests itself as an $\Raa$ value lower 
than 1 at large $\pt$.
Already at RHIC energies, the production of high $\pt$ 
(in the range 5-10~$\gev/c$) hadrons was found to be strongly suppressed 
(by a factor five) in the most central collisions~\cite{RAArhic}.

The higher centre-of-mass energy attained at the LHC allows extension of the 
$\pt$ range where the $\Raa$ is measured, thus providing important constraints
to the particle energy loss models.
Moreover, the more abundant production of charm and beauty enables 
higher precision measurements of the energy loss for heavy quarks, thus 
providing a further benchmark for theoretical models.
Radiative energy loss models predict that quarks lose less
energy than gluons (that have a larger colour charge) and that the amount of
radiated energy decreases with increasing quark mass.
Hence, a hierarchy in the values of the nuclear modification factor is 
anticipated, namely the $\Raa$ of B mesons should be larger than that of
D mesons that should in turn be larger than that of light-flavour hadrons
(e.g. pions), which mostly originate from gluon fragmentation.

\begin{figure}[tb!]
\centering
\includegraphics[width=0.42\textwidth]{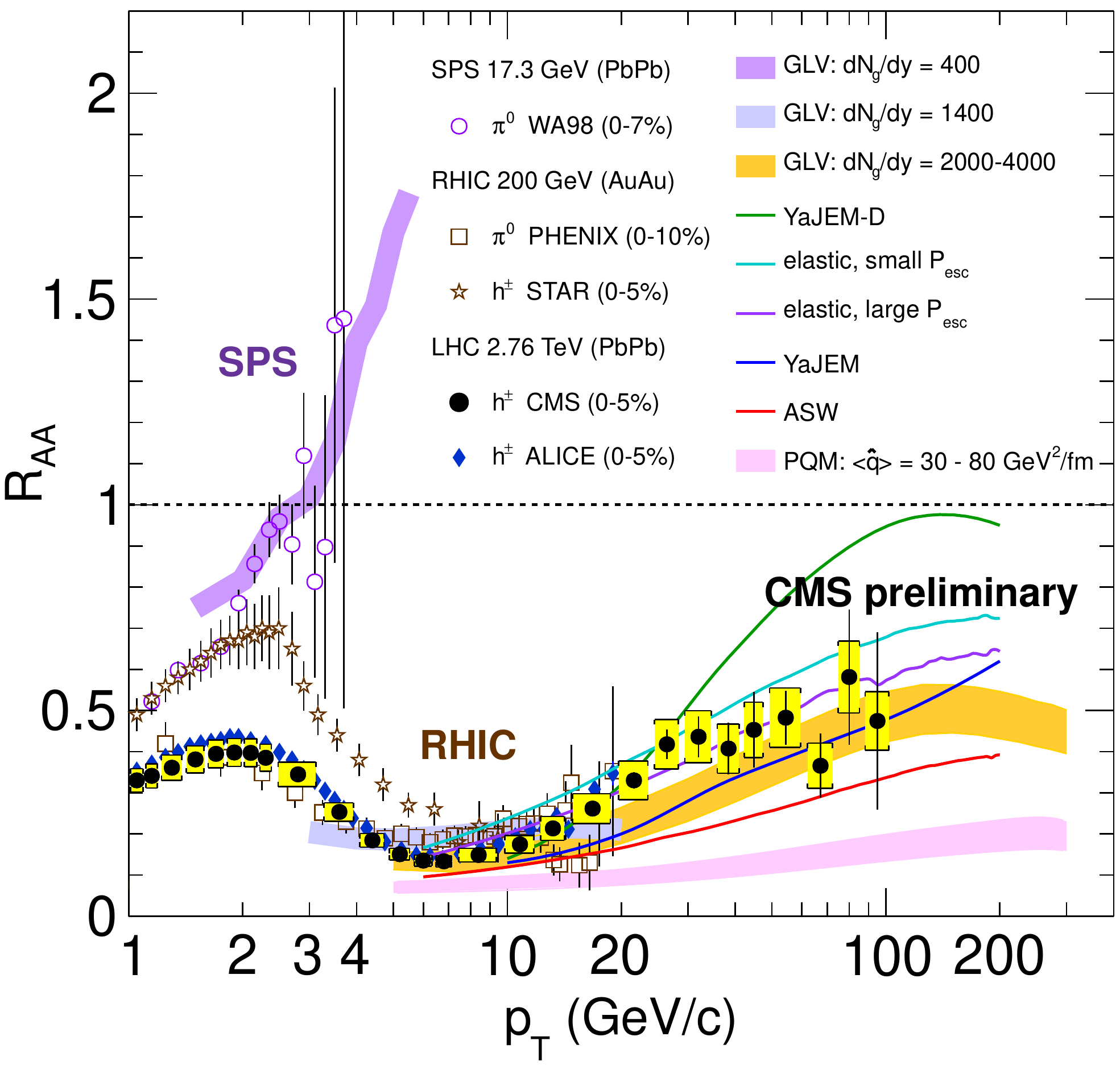}
\caption{Charged hadron nuclear modification factor at the LHC compared
to measurements at lower energies (RHIC and SPS). Taken from~\cite{CMSraa}.}
\label{fig:RAA}
\end{figure}

The $R_{AA}$ of unidentified charged particles has been measured by 
ALICE~\cite{ALICEraa} and CMS~\cite{CMSraa}. 
In particular, the CMS $\Raa$ (which extends up to $\pt=100~\gev/c$) 
is shown in Fig.~\ref{fig:RAA} for the 0--5\% centrality class compared to 
the ALICE measurement and to results from experiments at lower energies.
The $\Raa$ is smaller at the LHC, suggesting a larger energy loss relative
to RHIC energies and indicating that the density of the medium created in the 
collision increases with the increase of $\sqrts$.
It should also be considered that the fraction of hadrons originating from 
gluon jets increases with increasing centre-of-mass energy and, in radiative 
energy loss models, this is expected to give rise to a lower $\Raa$ because 
gluons lose more energy than quarks while traversing the QGP.
The $\Raa$ presents a minimum at $\pt\approx6-7~\gev/c$ and then increases 
slowly up to about 40~$\gev/c$. For $\pt$ above 40~$\gev/c$ the $\Raa$ seems
to level off at a value of approximately 0.5. Also at momenta of 100~$\gev/c$ 
a significant suppression of charged hadron yield is still present.

Medium-blind probes, such as photons and $Z^0$ bosons, can be used to test
that the initial production of hard probes follows the expected binary scaling, 
so as to separate initial and final state effects in the observed $\Raa$.
High energy (prompt) photons are produced directly from the hard scattering of 
two partons.
In nuclear collisions, they traverse the produced medium without suffering from
strong interaction, thus providing a direct test of pQCD production and 
initial state effects (such as nuclear modification of the parton distribution 
functions).
From the experimental point of view, the measurement of prompt photons is
challenging because of the huge background from $\pi^0$ and $\eta$ decays.
CMS has performed a measurement of isolated 
photon production in $\PbPb$ collisions as a function of 
centrality~\cite{CMSphotons}.
The measured $\Et$ spectra of isolated photons in $\PbPb$ are compared with
a theoretical (NLO pQCD) $\pp$ reference from JETPHOX 1.2.2~\cite{jetphox}
that provides a good description of the photon cross section in $\pp$ 
collisions at $\sqrt{s}=7~\tev$.
The resulting $\Raa$ is found to be compatible with unity within uncertainties.
This confirms that the initial production of hard probes follows the 
expected scaling of the production rate from $\pp$ to AA by the number 
of nucleon-nucleon collisions.
Z bosons have been measured in $\PbPb$ collisions by ATLAS~\cite{ATLASjpsi} 
and CMS~\cite{CMSz} via their decay into $\mu^+\mu^-$.
The measured Z$^0$ yield scaled by the number of nucleon-nucleon collisions
is found to be independent of centrality within uncertainties.
Furthermore, the Z$^0$ yield in $\PbPb$ results to be compatible
within uncertainties with the theoretical next-to-leading order pQCD 
$\pp$ cross sections scaled with the number of binary collisions.

\begin{figure}[tb!]
\centering
\includegraphics[width=0.42\textwidth]{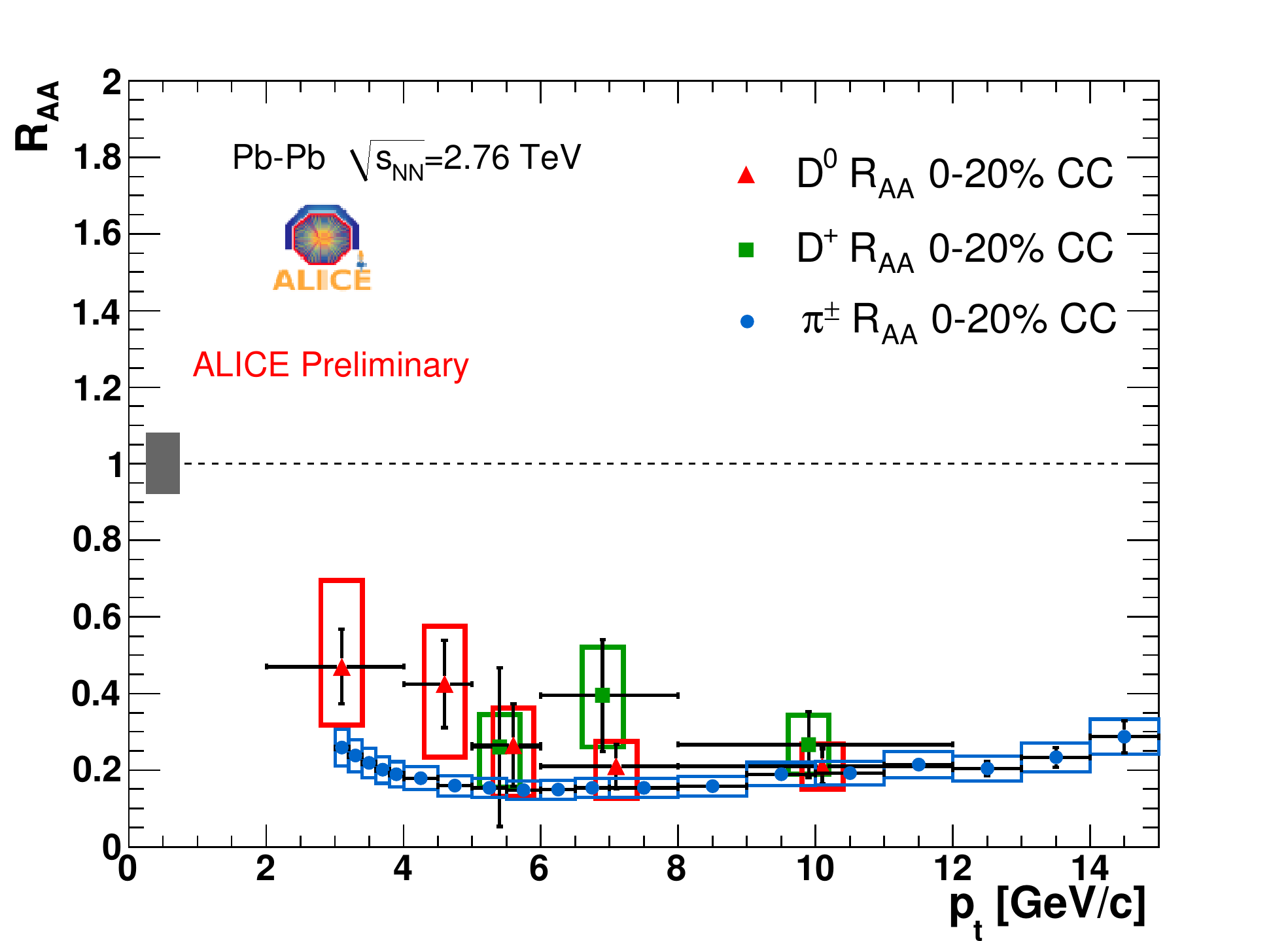}
\caption{Nuclear modification factor for prompt $\Dzero$ and $\Dplus$ mesons
in the 20\% most central collisions compared with the charged pions. 
Statistical (bars), systematic (empty boxes), and normalization (full box) 
uncertainties are shown. Taken from~\cite{DaineseQM}.}
\label{fig:Dmesons}
\end{figure}

As mentioned above, to provide further insight into the energy loss 
mechanisms, it is interesting to measure the nuclear modification factors 
for identified particles, in particular for heavy-flavoured hadrons.
The suppression of open charm and open beauty has been measured by ALICE 
with three different techniques: exclusive reconstruction of $\Dzero$
and $\Dplus$ hadronic decays at mid-rapidity, single electrons after subtraction
of a cocktail of background sources at mid-rapidity, and single muons at 
forward rapidity~\cite{DaineseQM}.
The measurement of prompt $\Dzero$ and $\Dplus$ $\Raa$ is shown in the left 
panel in Fig.~\ref{fig:Dmesons}.
A strong suppression is observed, reaching a factor 4-5 for
$\pt > 5~\gev/c$.
At high $\pt$ the suppression is similar to the one observed for charged 
pions, while at low $\pt$ there seems to be an indication for 
$\Raa$(D)$~>~\Raa$($\pi^\pm$).

As mentioned in section~\ref{sec:quarkonia}, CMS has measured the $\Raa$ of 
displaced $\jpsi$, i.e. coming from B meson decays, and reported a value of 
$\Raa=0.36 \pm 0.08 \pm 0.03$ for the 20\% more central collisions.
Also in this case, $\jpsi$ mesons are measured for $\pt>6.5~\gev/c$.
The physical mechanism behind this suppression is the $b$ quark energy loss.
The larger $\Raa$ of $\jpsi$ from $b$ with respect to that 
measured for prompt D mesons is a first indication for the $b$ quark to lose
less energy that the $c$ quark, as anticipated by radiative energy loss models.

\subsection{Jet quenching}

In-medium QCD energy loss is also predicted to affect fully reconstructed jets, 
due to ``jet quenching''.
The study of jet modification in heavy ion collisions is predicted to 
be a powerful tool to probe the properties of the QGP.
In particular, in-medium parton energy loss can significantly alter the 
observed energy balance between the most energetic (``leading'') and the 
second most energetic (``sub-leading'') jets in the event.
It is also interesting to measure the azimuthal angle between the two jets 
and verify if significant deviations with respect to back-to-back emission 
($\Delta\varphi_{\rm dijet}\approx\pi$) are observed.

Jet reconstruction in the high-multiplicity environment of high energy 
heavy-ion collisions is a challenging task that was pioneered in Au-Au 
reactions at RHIC.
In particular, jet measurements have been carried on in STAR: inclusive 
jet cross-section, di-jet coincidence rate and hadron-jet correlations. 
They  all converge toward a picture of broadening and softening of the jet 
fragmentation~\cite{STARjets}. 

At the LHC, ATLAS and CMS reported results with fully reconstructed jets, 
using the energies measured in $\eta,\varphi$ cells with the calorimeters 
and analyzed with different algorithms (anti-kt, iterative cone) for jet 
reconstruction as well as with different techniques for the subtraction of the 
background from the underlying event.

ATLAS~\cite{ATLASdijet} reported shortly after the $\PbPb$ run the observation
of a significant imbalance between the measured transverse energies of di-jets 
in opposite hemispheres.
The jet energy imbalance is quantified by the asymmetry $A_J$ defined as:
\begin{equation}
A_J=\frac{E_{T1}-E_{T2}}{E_{T1}+E_{T2}}
\end{equation}
where $E_{T1}$ is the transverse energy of the leading jet and $E_{T2}$ the
one of the most energetic jet in the opposite hemisphere.
$A_J$ is positive by construction.
It is required that $E_{T1}>100~\gev$ and $E_{T2}>25~\gev$ and therefore 
di-jets in which the sub-leading jet is below the $25~\gev$ threshold are not 
included in the $A_J$ calculation.

The measured distributions of $A_J$ in peripheral $\PbPb$ collisions is
similar to that observed in \pp\ events as well as to that expected from
Monte Carlo simulations without in-medium energy loss.
For more central collisions, the $A_J$ distribution broadens, does not
show a peak at $A_J=0$ and its mean is shifted to higher values.
In particular, for the most central events a peak is visible at $A_J\approx0.5$.
The different characteristics of the $A_J$ distribution in central $\PbPb$ 
collisions with respect to \pp\ indicate an increased rate of highly 
asymmetric di-jet events.
This observation is consistent with a degradation of the parton energy
while traversing the medium produced in $\PbPb$ collisions.
The azimuthal separation between the two jets 
has also been studied and shows that the two jets are essentially 
back-to-back in all centrality bins with no additional azimuthal broadening 
relative to that expected from QCD effects in vacuum.

Di-jet imbalance studies have also been performed by the 
CMS collaboration~\cite{CMSdijet} in a different energy range ($\pt$ 
thresholds at 120 and 50$~\gev/c$ for the leading and sub-leading jets 
respectively) and lead to the same conclusion about the presence of
a significant excess of imbalanced di-jets in central collisions.
This modification of the jet momentum balance implies a corresponding
modification in the distribution of the jet fragmentation products which
can be addressed by studying track-jet correlations.
In particular, the missing jet energy can be either transported out of the 
cone area used to define the jet, or it can be carried by low-momentum 
particles which are not measured in the calorimeter jets.
The overall momentum balance in the di-jet events was studied using the 
projection of the missing $\pt$ of reconstructed charged tracks onto the
leading jet axis:
\begin{equation}
\displaystyle{\not} p_{\mathrm{T}}^{\parallel} = 
\sum_{\rm i}{ -p_{\mathrm{T}}^{\rm i}\cos{(\phi_{\rm i}-\phi_{\rm Leading\ Jet})}},
\label{misspt}
\end{equation}
The results of this analysis show that, for charged-particle jets, if tracks 
down to $\pt=0.5~\gev/c$
are considered, the momentum balance is indeed recovered within uncertainties
also for the most central collisions.
In particular, a large negative contribution to $\missptave$
(i.e. in the direction of the leading jet) comes from particles with 
$\pt>8~\gev/c$ that is balanced by the contribution from lower $\pt$ 
particles, with a large fraction of the balancing momentum carried by tracks
with $\pt<2~\gev/c$.
Further insight is obtained by studying $\missptave$ separately
for tracks inside and outside cones of size 
$\Delta R=\sqrt{\Delta\varphi^2+\Delta\eta^2}=0.8$ around the leading 
and sub-leading jets.
The results show that a large part of the energy needed for balancing the 
jet momenta is carried by soft particles ($\pt < 2~\gev/c$) radiated at large 
angles with respect to the jet axes ($\Delta R > 0.8$)~\cite{CMSdijet}.

The observed large fraction of highly imbalanced di-jets in central $\PbPb$ 
collisions and the corresponding softening and widening of the
fragmentation pattern of the sub-leading jet are
consistent with a high degree of jet quenching in the 
produced medium.

First studies of fragmentation functions in $\PbPb$ collisions have been 
performed by ATLAS~\cite{ATLASff} and CMS~\cite{CMSff}.
The results show that the hard component of jet fragmentation functions in 
heavy-ion collisions resemble those of PYTHIA di-jet events (with partons
fragmenting in the vacuum) for both leading and sub-leading jets, 
in different centrality bins and different di-jet imbalance ($A_J$) bins.

\section{Summary and conclusions}
\label{sec:conclusions}

The first $\PbPb$ run at the LHC enabled the study of heavy ion physics 
at a center of mass energy about 14 times higher than at the largest
$\sqrts$ attained at RHIC.
ALICE, ATLAS and CMS experiments collected data during this first heavy ion run 
demonstrating excellent performance and complementary capabilities, that 
allowed (together with the impressive performance of the accelerator) 
a large variety of high quality results for all the experimental
observables that have been studied.
The overall picture that emerges from the ``soft'' (low $\pt$) sector 
is that the medium produced in $\PbPb$ collisions at the LHC is strongly 
interacting, has exceptionally low viscosity and can be well described by 
hydrodynamics.
In general, the results from soft physics observables show a smooth evolution 
from RHIC to LHC.
By comparing the high precision measurements at the LHC with model predictions 
and with the results obtained at lower energies at the SPS and at RHIC, it 
will be possible to obtain a deeper insight into the properties of the QGP.
Furthermore, with the first heavy ion run, the LHC experiments started to 
exploit the abundance of high $\pt$ and large mass probes  
which is a unique and novel aspect of heavy-ion collisions at LHC energies.
From the hard physics sector we expect to find answers to some of the issues
that have not been addressed completely by the experiments at lower 
collision energies, such as the J/$\psi$ (and quarkonia) suppression, 
the prediction of a different energy loss of beauty, charm and light hadrons 
and the in-medium modification of jet fragmentation properties.

\bigskip % extra skip inserted
\begin{acknowledgments}
The author wishes to thank R. Arnaldi, E. Bruna, A. Dainese, P. Giubellino,
R. Granier de Cassagnac, M. Masera and L. Ramello for useful discussions and 
suggestions during the
preparation of the conference talk and of these proceedings.
\end{acknowledgments}

\end{document}